\documentclass[aps,prd,twocolumn,superscriptaddress,longbibliography]{revtex4-2}  

\usepackage{graphicx}  
\usepackage[caption=false]{subfig}
\usepackage{dcolumn}   
\usepackage{bm}        

\usepackage{amssymb}   

\usepackage{physics}
\usepackage{xcolor}
\usepackage[normalem]{ulem}

\usepackage{url}\usepackage{hyperref}
\usepackage{bibunits}

\hypersetup{colorlinks,linkcolor={blue!55!black},citecolor={red!50!black},urlcolor={blue!45!black},breaklinks=true}

\def\6{{\langle}}
\def\9{{\rangle}}

\usepackage{dutchcal}

\usepackage{scalerel}
\usepackage{tikz}
\usetikzlibrary{svg.path}
\definecolor{orcidlogocol}{HTML}{A6CE39}
\tikzset{
	orcidlogo/.pic={
		\fill[orcidlogocol] svg{M256,128c0,70.7-57.3,128-128,128C57.3,256,0,198.7,0,128C0,57.3,57.3,0,128,0C198.7,0,256,57.3,256,128z};
		\fill[white] svg{M86.3,186.2H70.9V79.1h15.4v48.4V186.2z}
		svg{M108.9,79.1h41.6c39.6,0,57,28.3,57,53.6c0,27.5-21.5,53.6-56.8,53.6h-41.8V79.1z M124.3,172.4h24.5c34.9,0,42.9-26.5,42.9-39.7c0-21.5-13.7-39.7-43.7-39.7h-23.7V172.4z}
		svg{M88.7,56.8c0,5.5-4.5,10.1-10.1,10.1c-5.6,0-10.1-4.6-10.1-10.1c0-5.6,4.5-10.1,10.1-10.1C84.2,46.7,88.7,51.3,88.7,56.8z};
	}
}

\newcommand\orcidlink[1]{\href{https://orcid.org/#1}{\mbox{\scalerel*{
				\begin{tikzpicture}[yscale=-1,transform shape]
					\pic{orcidlogo};
			\end{tikzpicture}}{X}}}}

\begin{document}

\preprint{APS/123-QED}

\title{Can a quantum circuit detect the Unruh effect?
}

\author{Pravin Kumar Dahal\,\orcidlink{0000-0003-3082-7853}}
\email{pravin.dahal@csiro.au}
\affiliation{Commonwealth Scientific and Industrial Research Organisation (CSIRO), Clayton, Victoria 3168, Australia}

\author{Timothy C. Ralph} 
\affiliation{School of Mathematics and Physics, The University of Queensland, St Lucia, Australia}

\author{William J. Munro\,\orcidlink{0000-0003-1835-2250}} 
\affiliation{Okinawa Institute of Science and Technology Graduate University, Onna-son, Okinawa 904-0495, Japan}

\author{Arkady Fedorov} 
\affiliation{School of Mathematics and Physics, The University of Queensland, St Lucia, Australia}

\author{James Q. Quach} 
\email{james.quach@csiro.au}
\affiliation{Commonwealth Scientific and Industrial Research Organisation (CSIRO), Clayton, Victoria 3168, Australia}

\begin{abstract}

The Unruh effect predicts that an accelerating observer perceives the Minkowski vacuum as a thermal bath, yet direct detection remains experimentally inaccessible. Its timelike counterpart, arising from the entanglement of massless fields between the future and past light cones, offers a more feasible route but requires a detector whose transition frequency follows a specific conformal-time scaling. We propose and analyze a practical implementation of such a detector using superconducting fluxonium circuits, which naturally provide two quasi-degenerate ground states and a tunable excited state, forming an effective $\Lambda$-system. By modulating the excited-state transition frequency in Minkowski time, the detector accumulates a geometric phase associated with the timelike Unruh effect. Open-system simulations predict $\sim 10\%$ shift in the ground-state population within $530$ ns, representing a three-order-of-magnitude sensitivity enhancement over two-level Unruh–DeWitt detectors. These results establish a realistic quantum-circuit platform for experimentally probing the timelike Unruh effect and, more broadly, for testing fundamental nature of quantum fields using engineered quantum systems.

\end{abstract}

\maketitle

\begin{bibunit}[apsrev4-2]

Quantum field theory in curved spacetime~\cite{birrell1982,parker2009} emerged from efforts to quantize gravity and has yielded deep insights into the structure of spacetime. A key result is the Unruh effect~\cite{birrell1982,unruh1976,crispino2007,sciama1981,ginzburg1987}: an accelerating observer in Minkowski spacetime perceives the vacuum as a thermal bath, with the temperature proportional to the acceleration. A uniformly accelerating observer in Minkowski spacetime can be modeled as one at rest in Rindler spacetime. Rindler coordinates divide Minkowski spacetime into four wedges, with modes of a quantum field in the left and right Rindler wedges being entangled. Because the uniformly accelerating observer is restricted to a single wedge, tracing out the inaccessible modes produces the Unruh effect. A similar entanglement between the future and past (FP) light cones, for the case of massless fields, gives rise to timelike Unruh effect~\cite{olson2010,higuchi2017}; an observer confined to either the past or future light cones likewise perceives the Minkowski vacuum as a thermal state, known as the timelike Unruh effect.

The standard model for detecting the Unruh effect is the Unruh–DeWitt detector, a two-level point monopole~\cite{dewitt1979,unruh1976}. These detectors rely on energy exchange with the field, but such interactions may be too weak to register measurable excitations in ultraweak fields. For example, achieving a response equivalent to a 1 K thermal bath requires accelerations of $10^{20}\mathrm{m/s}^2$~\cite{davies1974}. To address this limitation, alternative detectors that avoid direct energy exchange with the field have been proposed. One approach~\cite{martin-martinez2010,hu2012} measures the geometric (Berry) phase acquired by a detector coupled to a quantum field, with the field’s temperature encoded in the accumulated phase\cite{pancharatnam1956,sjqvist2008,berry1984}.

Detecting the timelike Unruh effect, which requires scaling the energy levels of a detector, should be more experimentally accessible than observing the standard Unruh effect~\cite{olson2010}. A detector designed to probe this effect through its geometric phase response is therefore promising with current technology. Ref.~\cite{quach2021} proposed a $\Lambda$-detector, with two degenerate ground states and one excited state, to capture the geometric phase and infer the Unruh temperature. Here, we present a realistic superconducting-circuit architecture, based on a flux-tunable split junction and a superinductive chain, designed for experimental detection of the timelike Unruh effect. A recent experiment~\cite{luo2025} reported observing this effect using a two-level detector implemented in a quantum system of trapped ion. In contrast, our superconducting-qubit architecture has the ability to capture the geometric phase, offering roughly three orders of magnitude higher sensitivity than conventional two-level detectors and enabling broader applications in quantum metrology.

\textit{Fluxonium qubit as a $\Lambda$-detector--} We require a $\Lambda$-system exhibiting two key features that are not simultaneously realized in previous implementations~\cite{novikov2016,earnest2017}: (1) quasi-localized ground states that interact only via the excited state, and (2) tunable excited-state energy. The second feature is readily available in quantum circuits based on superconducting qubits. A typical superconducting qubit architecture consists of a small Josephson junction, which can be shunted by a capacitance and can also be connected in series with either multiple large-area junctions or an inductive shunt~\cite{krantz2019,kjaergaard2019}. This design framework encompasses transmons~\cite{koch2007}, flux qubits~\cite{orlando1999}, and their derivatives. Although superconducting $\Lambda$-systems have been demonstrated using cavity or resonator couplings~\cite{novikov2016}, these implementations generally lack doubly degenerate ground states, making them unsuitable for our purposes. 
A promising candidate that satisfies the first criterion is the fluxonium qubit—a subclass of flux qubits that supports quasi-localized states in separate potential wells. Transitions between these states are exponentially suppressed, yielding long coherence times that can extend into the millisecond range~\cite{manucharyan2009,nguyen2019}.

The basic fluxonium architecture includes a small Josephson junction shunted by a large superinductor $L_{JA}$ (constructed from an array of over 100 large-area junctions) and a capacitance $C_q$. 
Increasing the capacitance of the fluxonium enhances state localization, ensuring that the ground states couple only to the excited state and not to each other, thereby yielding the heavy fluxonium variant~\cite{earnest2017}. Although $\Lambda$-systems based on heavy fluxonium have been demonstrated~\cite{earnest2017}, these implementations lack the required tunability. To address this, we propose a new design featuring two key modifications: (1) replacing the single weak junction with a flux-tunable split junction, and (2) substituting the original superinductor with one engineered such that the inductive energy is greater than the charging energy~\cite{strickland2025,bell2012,masluk2012}.

\begin{figure}[!tbp]
    \centering
    \includegraphics[width=0.35\textwidth]{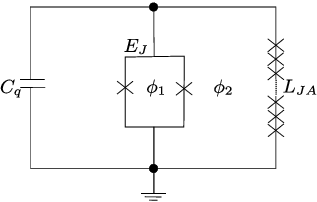}
    \caption{Circuit diagram of a tunable modified fluxonium circuit designed to operate as a $\Lambda$-system. Unlike earlier fluxonium designs, this circuit incorporates a flux-tunable split-junction and an inductive chain with $E_L>E_C$. The inductive chain is made up of series of large area Josephson junctions.}
    \label{fig:2}
\end{figure}

The circuit diagram incorporating these changes is shown in Fig.~\ref{fig:2}. Instead of the fixed Josephson energy $E_J$, the circuit Hamiltonian is defined by the variable Josephson energy of the split junction $E_J(\phi_1)$, as well as by two normalized external fluxes $\phi_1$ and $\phi_2$ piercing the split junction and the main loops, respectively:
\begin{equation}
    H_f= -4 E_C\frac{d^2}{d\phi^2}- E_J(\phi_1) \cos\left(\phi- \phi_J(\phi_{1})-\phi_2 \right)+\frac{1}{2} E_L\phi^2~, \label{Eq:ham1}
\end{equation}
where $E_C = e^2 / 2C_q$ is the charging energy, $E_L = (\Phi_0/2\pi)^2 /L_{JA}$ is the inductive energy ($\Phi_0 = h/2e$), $\phi$ is the superconducting phase across the junction and $\phi_J(\phi_1)= \tan^{-1} (d\tan\phi_1/2)$. Here,
\begin{equation}
    E_J(\phi_1)= E_{J\Sigma} \left|\cos\frac{\phi_1}{2}\right| \sqrt{1+ d^2 \tan^2\frac{\phi_1}{2}}~,
\end{equation}
$d= |E_{J1}- E_{J2}|/E_{J\Sigma}$ is the junction asymmetry and $E_{J\Sigma}= E_{J1}+ E_{J2}$ is the sum of the Josephson energies of the split junction. Like usual fluxonium devices, this circuit also allows for two types of transitions: interwell transitions between potential wells corresponding to different flux quanta in the loop formed by the junctions, and intrawell transitions between different levels within a well. An interwell transition is analogous to the transition of a flux qubit, but it is about $10^2-10^3$ times less sensitive to flux noise due to the large number of junctions in the fluxonium loop~\cite{lin2017}. As long as two adjacent wells are offset against each other, interwell transitions would have a vanishing overlap. Interwell transitions therefore are forbidden in the sense that any operator $O(\phi)$ would have exponentially small matrix elements for a sufficiently large ratio $E_J/E_C$~\cite{earnest2017}. This makes the metastable state very long-lived, with a lifetime extending into the millisecond regime.

\begin{figure*}[!tbp]
\centering

\subfloat[\label{2A}]{\includegraphics[width=5.77cm]{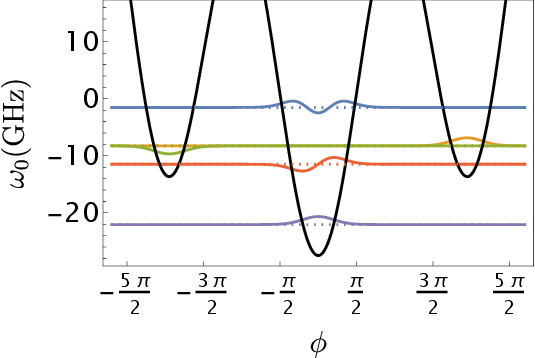}}\quad
\subfloat[\label{2B}]{\includegraphics[width=5.75cm]{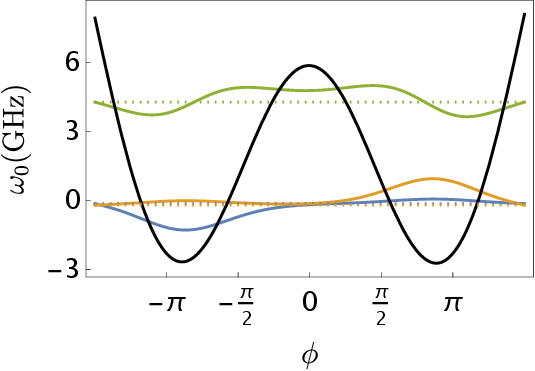}}\quad
\subfloat[\label{2C}]{\includegraphics[width=5.75cm]{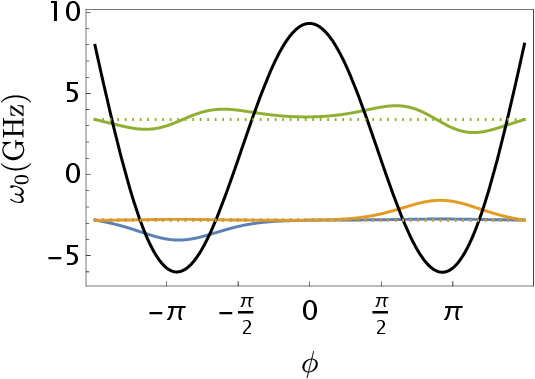}}
        
\caption{Potential energies (black curves), smallest eigenvalues (dotted lines), and corresponding eigenfunctions (solid curves excluding black) as functions of the total flux $\phi$ of the tunable modified fluxonium circuit of Fig.~\ref{fig:2} for: (a) $\phi_1=0, \phi_2=0$; (b) $\phi_1=2.50, \phi_2=0.88 \pi$;
and (c) $\phi_1=2.80, \phi_2= 0.78 \pi$. Here, $\phi_1$ and $\phi_2$ represent the external magnetic fluxes through the split junction and the main loops of the circuit, respectively. Fig.~(a) shows the circuit with multiple potential wells of varying depths, which depend on the external field. We tune the external field so that the depths of two adjacent wells become equal, as illustrated in Figs.~(b) and (c). In this regime, the transition dipole between the two lowest levels is negligible compared to the transition dipoles between these lowest levels and the excited state, thereby illustrating the circuit's operation as a $\Lambda$-system. We initialize the circuit by setting specific values for the external fluxes $(\phi_1,\phi_2)=(2.50, 0.88\pi)$, which correspond to an initial frequency in Minkowski time $\omega^i_M/2\pi= 4.4$ GHz, as illustrated in Fig.~(b). By continuously adjusting the external fluxes to $(\phi_1,\phi_2)=(2.80, 0.78\pi)$, we achieve a continuous tuning of the frequency to the final value $\omega^f_M/2\pi= 6.2$ GHz, as shown in Fig.~(c). These configurations provide the necessary tunability of the energy levels, which is essential for detecting the timelike Unruh effect. Outside these flux ranges, the circuit does not function as a $\Lambda$-system.}
\label{fig:3}
\end{figure*}

Near $\phi_1=0=\phi_2$, the low energy spectrum corresponds to intrawell transitions in the central potential well with transition dipoles similar to those of transmons (see Fig.~\ref{2A}). Instead, we work in the so-called ``sweet spot", where the leading order transition frequency vanishes because of the symmetry~\cite{nguyen2019}. In this case, the states $|g_1\rangle$ and $|g_2\rangle$ correspond to the tunnel splitting of the two-fold degenerate ground state. The excited state $|e\rangle$ is separated by a plasma gap (see Fig.~\ref{2A}). Higher states also exists and form an anharmonic spectrum with a rich selection rule structure.


Using the parameter values commonly adopted in the literature~\cite{krantz2019,lin2017}, the modified fluxonium circuit functions as a tunable $\Lambda$-system (the parameter values are given in Supplementary Information Sec.~V). Its energy levels can be tuned within the range $\omega^i_M/2\pi= 4.4$ GHz to $\omega^f_M/2\pi= 6.2$ GHz, where $\omega_M$ denotes the frequency in Minkowski time (see Figs.~\ref{2B} and \ref{2C}). As its name suggests, such a system has two degenerate ground states ($(|g_1\rangle, |g_2\rangle)$) and an excited state $(|e\rangle)$. We have verified numerically that the transition $|g_1\rangle\to |g_2\rangle$ has a vanishing transition dipole $\langle g_2|\phi|g_1\rangle$ and hence is a forbidden type. This ensures that the metastable state $|g_2\rangle$ is very long-lived and thus approximately degenerate to the ground state for our purpose. However, the transitions $|g_1\rangle\to |e\rangle$ and $|g_2\rangle\to |e\rangle$ have transition dipoles $\langle e|\phi|g_1\rangle$ and $\langle e|\phi|g_2\rangle$ of the order unity and are thus allowed type. Thus, the nearly degenerate ground states $|g_1\rangle$ and $|g_2\rangle$ interact only through $|e\rangle$, thereby justifying that $|g_1\rangle$, $|g_2\rangle$ and $|e\rangle$ behave as a $\Lambda$-system is a good approximation.

\textit{Evolution in future-past light cones.} The effective Hamiltonian of such a $\Lambda$-system interacting with an electromagnetic field, where the ground states only couple to the excited state with a transition frequency $\omega/2\pi$ and not with each other, can be written as $H= H_0+ H_I$, where $H_0= \hbar\omega|e\rangle\langle e|$ constitutes its free energy and
\begin{equation}
    H_I= -q \mathbf{r}(\eta)\cdot \mathbf{E}(\eta)|e\rangle \left(\langle g_1|+ \langle g_2|\right)+ H.c.
\end{equation}
is the detector-field interaction in worldline $(\eta,0)$. Here, $q \mathbf{r}(\eta)$ is the dipole moment operator of the detector, $\mathbf{E}(\eta)= - \partial\mathbf{A}(\eta)/\eta$ is the electric field strength with $\mathbf{A}(\eta)$ being the vector potential and $H.c.$ denotes the Hermitian conjugate. The time coordinate $\eta$ is related to Minkowski time by Eq.~\eqref{eq:ct6}. Let us redefine the basis $|\pm\rangle= \left(|g_1\rangle\pm |g_2\rangle\right)/\sqrt{2}$ and operators $\sigma_{ij}\equiv |i\rangle \langle j|$. In this basis, interaction Hamiltonian becomes $H_I= -q \mathbf{r}\cdot \mathbf{E} (\sigma_{e+} + \sigma_{e+}^\dagger )$.
It can be observed that the state $|-\rangle$ does not interact with the electric field and is thus termed a dark state. The interaction Hamiltonian thus reduces to that of a two-level system.

Decoherence of the state $|g_2\rangle$ could, in principle, induce population transfer between the $|-\rangle$ and $|+\rangle$ states, thereby causing the evolution of $|-\rangle$. However, the decoherence rate of $|g_2\rangle$ in a fluxonium device operating at the sweet spot is on the order of $10^4$~\cite{nguyen2019}. This rate is negligible compared to the longitudinal $\Gamma_1$ and transverse $\Gamma_2$ relaxation rates of the excited state $|e\rangle$ in fluxonium qubits, which are on the order of $10^6$. Consequently, the contribution of the decoherence of $|g_2\rangle$ to the evolution of the bright state can be safely ignored. Therefore, only the bright components of the $\Lambda$-system evolve, and its dynamics in a thermal bath can be conveniently expressed in the $\{|+\rangle,|-\rangle\}$ basis as (see Supplementary Information Sec.~III):
\begin{equation}
    \frac{\partial\rho}{\partial \eta}= -\frac{i}{\hbar}\left[H_0, \rho\right]+ {\cal P}(\omega) {\cal L}[\sigma_{e+}]- {\cal P}(-\omega) {\cal L}[\sigma_{e+}^\dagger]~, \label{eq:linde}
\end{equation}
where ${\cal L}[O]= O \rho O^\dagger - \{O^\dagger O, \rho\}/2$. Here, ${\cal P}(\omega)$ denotes the transition probability per unit proper time as the $\Lambda$-detector is tuned between $\omega^i_M$ and $\omega^f_M$. 

Tuning the $\Lambda$-detector between $\omega^i_M$ and $\omega^f_M$ is equivalent to restricting it either in the future or the past light cone, where it follows the worldline $(\eta, 0)$. A $\Lambda$-system can capture the geometric phase associated with the timelike Unruh effect when restricted to either the future or past Rindler light cone. To obtain the theoretical prediction for the geometric phase, we first need to calculate the transition probability per unit time ${\cal P}(\omega)$. We apply the following coordinate transformations to the Minkowski metric ($ds^2= -c^2 dt^2+ dx^2+ dy^2+ dz^2$)
\begin{equation}
    t= a^{-1} e^{a\eta} \cosh{a\zeta}~, \quad z= c a^{-1} e^{a\eta} \sinh{a\zeta}~, \label{eq:ct6}
\end{equation}
where $a$ is an arbitrary parameter with units of Hz and $c$ is the velocity of light. The resulting metric is valid on the future Rindler light cone
\begin{equation}
    ds^2= e^{2a\eta}(-d\eta^2+ d\zeta^2)+ dx^2+ dy^2~.
\end{equation}
Similarly, we can define coordinate transformations $(\bar\eta,\bar\zeta)$
\begin{equation}
    t= -a^{-1} e^{a\bar\eta} \cosh{a\bar\zeta}~, \quad z= -c a^{-1} e^{a\bar\eta} \sinh{a\bar\zeta}~,
\end{equation}
such that the resulting metric is valid only on the past Rindler light cone. The vacuum state corresponding to the Minkowski metric is denoted by the Minkowski vacuum $\ket{0_M}$, while that corresponding to the Rindler metric is the Rindler vacuum $\ket{0_R}$. As in the conventional Unruh effect, vacuums in the FP light cones are entangled for the case of massless fields, thereby giving rise to the timelike Unruh effect~\cite{olson2010,higuchi2017}. Consider the Schrodinger equation for an observer in the worldline $(\eta,0)$: $i\hbar \partial|\psi\rangle/\partial \eta= H|\psi\rangle$. In Minkowski coordinates, this equation becomes
\begin{equation}
i\hbar\frac{\partial|\psi\rangle}{\partial t}= \frac{H}{a t} |\psi\rangle~.
\end{equation}
This implies that a detector at rest in the FP coordinates corresponds to a detector with energy levels scaled inversely with time in Minkowski coordinates. When the detector is restricted to the future or past light cone, the Minkowski vacuum induces a thermal response with respect to the detector’s evolution in conformal-time, so that the bright sector of the $\Lambda$-detector behaves as if it were coupled to a thermal bath and can undergo transition to the excited states. The transition probability per unit proper time associated with this process is~\cite{olson2010,quach2021}
\begin{equation}
    \begin{aligned}
        {\cal P}(\omega)=& \frac{q^2}{\hbar^2} |\langle\omega|\mathbf{r}(0)|0\rangle|^2 \int_{-\infty}^\infty d(\Delta\eta) e^{i\omega \Delta\eta} e^{a(\eta+\eta')} G^+(\Delta\eta)\\
        =& \frac{\Gamma(\omega)}{2} \left(1+ \frac{a^2}{\omega^2}\right) \left(1+ \coth\frac{\pi\omega}{a}\right)~, \label{eq:drpt}
    \end{aligned}
\end{equation}
where $G^+_{ij}$ is the two point function of the electric field and $\Gamma(\omega)= \omega^3 q^2 |\langle\omega|\mathbf{r}(0)|0\rangle|^2/(2\pi\epsilon_0\hbar c^3)$ is the spontaneous emission rate (see Supplementary Information Sec.~{I} for derivation). We thus have the response function equivalent to that of a detector in a thermal bath of temperature $T= \hbar a/ 2\pi k_B$. 

The solution for the density matrix $\rho(\eta)$ is readily obtained by taking a general initial state (see Supplementary Information Sec.~III for details)
\begin{equation}
    |\psi(0)\rangle = \sin\frac{\vartheta}{2} \left(\cos\frac{\theta}{2} |e\rangle + \sin\frac{\theta}{2} |+\rangle\right) + \cos\frac{\vartheta}{2} |-\rangle~.
\end{equation}
Resulting solution allows us to identify subspaces of bright eigenstates as
\begin{equation}
    |p_i(\eta)\rangle = e^{i\alpha_e(\eta)} \sqrt{p_{i,e}(\eta)} |e\rangle + e^{i\alpha_g(\eta)} \sqrt{p_{i,+}(\eta)} |+\rangle~,
\end{equation}
where $\alpha(\eta)$ is the dynamical phase. As the detector evolves along the $(\eta,0)$ worldline, both bright and dark components accumulate dynamical phases. However, only bright components acquires an additional geometric phase $\beta(\eta)$ due to its cyclic interaction with the field. Consequently, bright and dark states of the system can be written as $e^{i(\beta(\eta)+\alpha_b(\eta))} \sum_{i=1,2} \sqrt{p_i(\eta)} |p_i(\eta)\rangle$ and $e^{i\alpha_g(\eta)}\sqrt{p_-}|-\rangle$, respectively~\cite{quach2021}.

\textit{Geometric phase and population change.} The geometric phase acquired by bright components can be computed using the relation~\cite{tong2004}
\begin{equation}
    \beta= \text{arg}\sum_{i=1}^N \sqrt{p_i(0)p_i(T)}\langle p_i(0)|p_i(T)\rangle e^{-\int_0^T d\eta \langle p_i(\eta)|\dot p_i(\eta)\rangle}~. \label{eq:gpr}
\end{equation}
An explicit expression for the total geometric phase acquired by the $\Lambda$-system during its evolution is obtained by substituting the bright states eigenvalues $p_i(\eta)$ into Eq.~\eqref{eq:gpr} (see Supplementary Information Sec.~II for details). 
The advantage of using the $\Lambda$-system is that its geometric phase can be determined by monitoring the ground state population $P_1(\eta)= \langle g_1|\rho(\eta)|g_1\rangle$. The expression for the ground state population as it evolves in the thermal bath is
\begin{multline}
    P_1(\eta)= \frac{1}{2}\left(1 -\sin^2\frac{\vartheta}{2} e^{-2 A \eta} f(\eta) \right)\\
    + \frac{1}{2} e^{-(A-B) \eta} \sin\vartheta \sin\frac{\theta}{2} \cos \beta(\eta)~, \label{eq:gsp}
\end{multline}
where
\begin{equation}
    f(\eta)= e^{-2 A \eta} \cos^2\frac{\theta}{2}- \left(\frac{B}{A}- 1\right) \sinh(2A \eta)~.
\end{equation}
Here, $A$ and $B$ are coefficients of the Kossakowski matrix of the system, whose values can be explicitly computed using the transition probabilities per unit time~\eqref{eq:drpt}
\begin{equation}
    A= \frac{1}{4} \left({\cal P}(\omega)+ {\cal P}(-\omega)\right), \quad B= \frac{1}{4} \left({\cal P}(\omega)- {\cal P}(-\omega)\right)~.
\end{equation}
To estimate the change in the ground state population due to the Unruh effect, we first substitute the known values of initial and final frequencies in Minkowski time, $\omega^i_M$ and $\omega^f_M$, into the relation for their ratio $\omega^f_M/\omega^i_M = e^{-a/\omega}$~\cite{Olson2011}. We obtain the corresponding acceleration in the $(\eta,0)$ worldline to be $a/\omega = 0.34$. The Minkowski time interval corresponding to $n$ quasicycle evolution interval in FP time is given as~\cite{Olson2011}
\begin{equation}
    t_c= \frac{\omega}{a \omega^i_M} \left(e^{n a/\omega}-1\right)~. \label{eq:timch}
\end{equation}
We want to restrict $t_c$ to the order of $1 \mu s$, which is below the timescale of superconducting qubits' lifetime. This relation implies that the Minkowski time interval of $1 \mu s$ corresponds to $n=25$, $25$-quasicycle evolutions in FP time. Conducting the experiment for $25$-quasicycles, without having to tune the energy gap outside the range given above, can be done by alternating between positive and negative $a$. This is because, the change in ground state population~\eqref{eq:gsp} is independent of the sign of $a$ when $\theta=\pi/2$. Thus, one switches the sign of $a$ every quasicycle and the frequency of the system oscillates between $\omega^i_M$ and $\omega^f_M$. Additionally, we take the spontaneous decay rate $\Gamma$ to be approximately $10^{-4} \omega$~\cite{nguyen2019}.


\textit{Experimental protocol--} To detect the timelike Unruh effect in a superconducting circuit, we implement a four-step protocol designed to isolate and quantify the population shift induced by the phenomenon. First, the circuit is initialized with external flux values $\phi_1 = \phi_{1i}$ and $\phi_2 = \phi_{2i}$. For the circuit in Fig.~\ref{fig:2}, with parameter values provided in Supplementary Information Sec. V, these initial fluxes are set to $\phi_{1i} = 2.50$ and $\phi_{2i} = 0.88\pi$, corresponding to the eigenspectrum in Fig.~\ref{2B}. The initial ground-state populations are distributed across the bright and dark subspaces to maximize the observable change in population induced by the Unruh effect. Specifically, the dark-state population is chosen as $p_- \approx 0.2$, which yields an approximately maximal population change.

Next, we measure the population change $P_1(0,\eta)- P_1(0,0)$ after $t_c=t_{cf}$ of evolution in Minkowski time. Here, $P_1(0,\eta)$ denotes the population obtained from measurements performed without modulation of the external fluxes, corresponding to zero Unruh temperature. For our circuit, $t_{cf} \approx 530$ ns.

The third step involves tuning the detector and measuring the corresponding population change $P_1(a,\eta)- P_1(a,0)$ as the detector evolves for $t_{cf}$ in Minkowski time. Tuning is achieved by oscillating the external fluxes between their initial values and final values $\phi_1=\phi_{1f}$ and $\phi_2=\phi_{2f}$, which causes the energy levels of the $\Lambda$-detector to oscillate between $\omega^i_M/2\pi$ and $\omega^f_M/2\pi$. For our circuit, the final external fluxes are $\phi_{1f} = 2.80$ and $\phi_{2f} = 0.78\pi$, corresponding to the eigenspectrum in Fig.~\ref{2C}. During this evolution, a geometric phase change is induced in the detector, which is captured by its ground state population $P_1(a,\eta)$, as given by Eq.~\eqref{eq:gsp}.

Since we are interested only in the population components arising from the Unruh effect, the final step is to determine the corresponding shift in the ground state population, defined as $\Delta_{P_1}(\eta)= \delta P_1(\eta)-\delta P_1(0)$, where $\delta P_1(\eta):= P_1(a,\eta)- P_1(0,\eta)$. This difference between measurements with and without modulation of the external fluxes isolates the contribution solely due to the Unruh effect.

\begin{figure}[!tbp]
    \centering
    \includegraphics[width=0.48\textwidth]{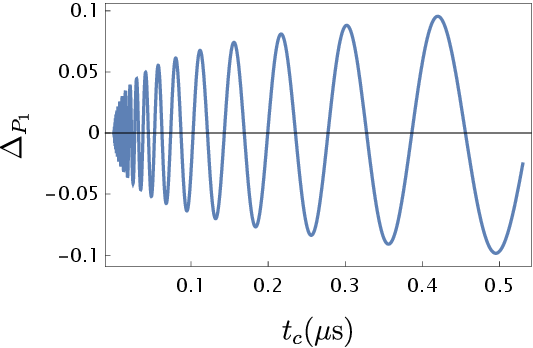}
    \caption{
    Evolution of the change in ground state population due to the Unruh effect during $530$ ns evolutions in Minkowski time (equivalent to approximately $25$-quasicycle evolutions in FP time). The plot shows that the ground state population shifts by approximately ten percent as a result of the timelike Unruh effect, thereby making the effect detectable. If there is no Unruh effect, there will be no change in population.}
    \label{fig:4}
\end{figure}

The energy levels of the $\Lambda$-detector oscillate between $\omega^i_M/2\pi= 4.4$ GHz and $\omega^f_M/2\pi= 6.2$ GHz. The profile of this tuning during the $530$ ns evolutions in Minkowski time can be obtained from Eq.~\eqref{eq:timch} (see Fig.~S1 for details). Over this interval, the energy levels undergo approximately $25$-quasicycles evolutions, where each quasicycle corresponds to an evolution in conformal time from $\eta=0$ to $\eta= 2\pi/\omega$. A plot of the shift in ground state population as a result of the Unruh effect is shown in Fig.~\ref{fig:4}. From this figure, we infer that over $25$-quasicycle evolutions in FP time (corresponding to approximately $530$ ns in Minkowski time), the ground state population shifts by about ten percent, ${\cal O}(\Delta_{P_1})\approx0.1$, due to the Unruh effect.

By monitoring the change in the ground state population of the $\Lambda$-detector, one therefore measures the Unruh temperature. Owing to its ability to accumulate geometric phase, the $\Lambda$-detector exhibits a sensitivity enhancement of three orders of magnitude compared to a two-level Unruh–DeWitt detector, which does not capture the geometric phase. For a two-level detector, the population change over $25$ quasicycles in FP time is ${\cal O}(\Delta_{P_1}) \approx 8.7 \times 10^{-4}$, which is obtained by setting the dark-state population to zero (corresponding to $\vartheta = \pi$) in Eq.~\eqref{eq:gsp}. Hence, geometric-phase accumulation in a $\Lambda$-detector significantly enhances sensitivity to the Unruh effect, enabling measurable population shifts within experimentally accessible timescales. By capturing the geometric phase, the $\Lambda$-detector exploits the phase offset between energy-tuned and untuned evolutions, thereby amplifying the magnitude of the difference in their population.

\textit{Conclusion--} We have designed a new circuit architecture for the tunable $\Lambda$-system by incorporating a flux-tunable split junction and a large superinductor, enabling the accumulation of geometric phase change induced by the timelike Unruh effect. Its advantage is the ability to tune the energy levels of the detector to make the effect observable under experimentally feasible conditions. The timelike Unruh effect can be detected by monitoring the ground state population of the $\Lambda$-system. Our numerical estimates indicate that the ground state population changes significantly due to the Unruh effect, making it a viable method for experimental detection. Besides detecting Unruh temperature, our quantum circuit has other potential applications, such as quantum thermometry and quantum non-demolition measurements. Thus, even a simple quantum circuit like the one discussed here has the potential to provide further insights into the fundamental nature of quantum fields and spacetime.

\section*{Acknowledgements}

PKD would like to thank Yaniv Kurman, Kieran Hymas and Alexander Nguyen for useful discussions. The work of PKD is supported by CSIRO Early Research Career postgraduate fellowship.

\putbib[origin1]
\end{bibunit}

\clearpage
\begin{bibunit}[apsrev4-2] \makeatletter \renewcommand{\bibnumfmt}[1]{[S#1]} \renewcommand{\citenumfont}[1]{S#1} \makeatother 
\widetext 
\begin{center} \textbf{\large Supplementary Information: Can a quantum circuit detect the Unruh effect?} \end{center}

\twocolumngrid

\setcounter{equation}{0}
\setcounter{figure}{0}
\setcounter{table}{0}
\setcounter{page}{1}
\makeatletter
\renewcommand{\theequation}{S\arabic{equation}}
\renewcommand{\thefigure}{S\arabic{figure}}
\renewcommand{\thetable}{S\arabic{table}}

\section{Timelike Unruh effect} \label{s2}

We consider the Unruh-Dewitt detector, which is the simplest particle detector model. The detector-field interaction of such a detector in worldline $(\eta,0)$ is described by the interaction Lagrangian $(e^{a\eta}) (-q \mathbf{r}(\eta)\cdot \mathbf{E}(\eta))$, where $q \mathbf{r}(\eta)$ is the dipole moment operator of the detector and $\mathbf{E}(\eta)= - \partial\mathbf{A}(\eta)/\eta$ is the electric field strength with $\mathbf{A}(\eta)$ being the vector potential. Here, we should note that the interaction Lagrangian differs from the conventional case by an additional factor of $e^{a\eta}$~\cite{olson2011}. This detector will undergo a transition from the ground state to an excited state as the field undergoes a transition from the vacuum $|0_M\rangle$ to the excited state $|0_R\rangle$. In the regime of our approximation, this transition amplitude is given by the first-order perturbation theory
\begin{equation}
    {\cal A}= -\frac{i q}{\hbar} \langle\omega,0_R|\int_{-\infty}^\infty e^{a\eta} \mathbf{r}(\eta)\cdot \mathbf{E}(\eta) d\eta |0\rangle~.
\end{equation}
Using the equation for the time evolution of $\mathbf{r}(\eta)$, we obtain
\begin{equation}
    \mathbf{r}(\eta)= e^{i H_0\eta} \mathbf{r}(0) e^{-i H_0\eta}~,
\end{equation}
where $H_0|\omega\rangle= \hbar \omega|\omega\rangle$. The transition probability to all possible $|0_R\rangle$ and $\omega$ can be obtained by squaring the modulus of the amplitude and summing over the complete set $|0_R\rangle$
\begin{equation}
    {\cal P}(\omega)= \frac{q^2}{\hbar^2} \sum_{i,j=1}^3 \langle\omega|r_i(0)|0\rangle \langle 0|r_j(0)|\omega\rangle {\cal F}_{ij} (\omega)~, \label{eq:tpb}
\end{equation}
where
\begin{equation}
    {\cal F}_{ij}(\omega)= \int_{-\infty}^\infty d\eta \int_{-\infty}^\infty d\eta' e^{i\omega(\eta-\eta')} e^{a(\eta+\eta')} G^+_{ij}(\eta,\eta')~, 
\end{equation}
is the detector's response function and $G^+_{ij}$ is the two point function of the electric field. ${\cal F}_{ij}(\omega)$ represents the bath of particles that the detector effectively experiences as a result of its motion. As this transition amplitude is computed for an infinite time integral, the transition probability will diverge if the number of quanta absorbed by a detector per unit time is non-zero~\cite{birrell1982}. To deal with this circumstance, we instead consider the transition probability per unit proper time
\begin{multline}
    \frac{q^2}{\hbar^2} \sum_{i,j=1}^3 \langle\omega|r_i(0)|0\rangle \langle 0|r_j(0)|\omega\rangle\\
    \int_{-\infty}^\infty d(\Delta\eta) e^{i\omega \Delta\eta} e^{a(\eta+\eta')} G^+_{ij}(\Delta\eta)~.
\end{multline}
Now, the two-point function for the electric field can be calculated from the regularized positive frequency Wightman function for the massless scalar field as
\begin{multline}
    G^+_{ij}(x,x')= -\frac{\hbar}{4\pi^2\epsilon_0 c}\\
    \left(\partial_t\partial_{t'}\delta_{ij}- \partial_i \partial_{j'} \right) \frac{1}{c^2 (t-t'-i\epsilon)^2- (\mathbf{x}-\mathbf{x'})^2}~,
\end{multline}
where $\epsilon_0$ is the permittivity of vacuum and $\delta_{ij}$ is the Kronecker delta. The functional form of $e^{a(\eta+\eta')}G^+_{ij}(\eta,\eta')$ for two points on the $(\eta,0)$ worldline
\begin{equation}
    t= a^{-1} e^{a\eta}~, \quad z=0~,
\end{equation}
is identical to that of $G^+_{ij}(\eta,\eta')$ for two points on the accelerated trajectory
\begin{equation}
    t= a^{-1} \sinh a\tau~, \quad z= c a^{-1} \cosh a\tau~,
\end{equation}
up to a rescaling of $\epsilon$. This can be seen through a straightforward coordinate substitution, noting in particular that for the $(\eta,0)$ worldline
\begin{multline}
    G^+(\eta,\eta')= -\frac{\hbar}{4\pi^2\epsilon_0 c^3} \left(\partial_t\partial_{t'}\delta_{ij}- \partial_i \partial_{j'} \right)\frac{1}{(t-t'-i\epsilon)^2}\\
    = \frac{3\hbar}{32\pi^2\epsilon_0 c^3} \frac{a^4 e^{-2 a(\eta+\eta')}}{\sinh^4\left(\frac{a}{2}(\eta-\eta')\right)}~. \label{eq:2pf}
\end{multline}
Hence, in correspondence to the Unruh effect, a detector in the $(\eta,0)$ worldline will respond at a temperature $T= \hbar a/ 2\pi k_B$. Thus, the entanglement of vacuums in the future and past Rindler light cones gives rise to the timelike Unruh effect. The temperature associated with the timelike Unruh effect is $T= \hbar a/ 2\pi k_B$.

Substituting Eq.~\eqref{eq:2pf} into Eq.~\eqref{eq:tpb}, we obtain the transition probability per unit time
\begin{equation}
    \begin{aligned}
        {\cal P}(\omega)=& \frac{q^2}{\hbar^2} |\langle\omega|\mathbf{r}(0)|0\rangle|^2 \int_{-\infty}^\infty d(\Delta\eta) e^{i\omega \Delta\eta} e^{a(\eta+\eta')} G^+(\Delta\eta)\\
        =& \frac{\Gamma(\omega)}{2} \left(1+ \frac{a^2}{\omega^2}\right) \left(1+ \coth\frac{\pi\omega}{a}\right)~, \label{eq:drptS}
    \end{aligned}
\end{equation}
where $\Gamma(\omega)= \omega^3 q^2 |\langle\omega|\mathbf{r}(0)|0\rangle|^2/(2\pi\epsilon_0\hbar c^3)$ is the spontaneous emission rate. We have thus recovered a response function of a detector in a thermal bath of temperature $T= \hbar a/ 2\pi k_B$.

It should be noted that the factor $e^{a\eta}$ appearing in the interaction Hamiltonian is a kinematic conformal factor associated with the transformation from Minkowski to future/past light-cone time; it does not play a role in an independent enhancement of the detector–field coupling. For a trajectory in the future/past light-cone, the Wightman function carries the compensating factor
\begin{equation}
    G^+(\eta,\eta')\propto e^{-a(\eta+\eta')}\tilde G^+(\eta-\eta')~,
\end{equation}
where $\tilde G^+(\eta-\eta')$ has the same functional form as the two-point correlator along a uniformly accelerated trajectory. Thus, the factor $e^{a\eta}$ does not induce an uncontrolled growth of the perturbative expansion; rather, the transition probability, to all orders, takes the same functional form as in the uniformly accelerated Unruh case because of it. This is reflected in Eq.~\eqref{eq:drptS}, where the resulting transition rate is of order $\Gamma(\omega)$ up to an ${\cal O}(1)$ prefactor.

\section{Geometric phase in $\Lambda$-detector} \label{s3}

The interaction Hamiltonian of a $\Lambda$-system reduces to that of a two-level system. Its bright eigenstates constitutes two subspaces $|p_1(\eta)\rangle$ and $|p_2(\eta)\rangle$, which can be defined generically as
\begin{equation}
    |p_i(\eta)\rangle= e^{i\alpha_e(\eta)}\sqrt{p_{i,e}(\eta)} |e\rangle+ e^{i\alpha_g(\eta)} \sqrt{p_{i,+}(\eta)} |+\rangle~, \label{eq:ssp}
\end{equation}
where $\alpha(\eta)$ is the dynamical phase. We can thus write an arbitrary bright state as a linear combination of these two subspaces
\begin{equation}
    |\psi_b(\eta)\rangle= \frac{\sqrt{p_1(\eta)} |p_1(\eta)\rangle+ \sqrt{p_2(\eta)} |p_2(\eta)\rangle}{\sqrt{p_1(\eta)+p_2(\eta)}}~. \label{eq:bsp}
\end{equation}
Bright and dark states of the system evolves in the field as $e^{i(\beta(\eta)+\alpha_b(\eta)+\varphi)}\sqrt{p_1(\eta)+p_2(\eta)}|\psi_b(\eta)\rangle$ and $e^{i\alpha_g(\eta)}\sqrt{p_-}|-\rangle$ respectively~\cite{quach2021}.


For a mixed state undergoing nonunitary quasicyclic evolution, the geometric phase can be calculated using the relation
\begin{equation}
    \beta= \text{arg}\sum_{i=1}^N \sqrt{p_i(0)p_i(T)}\langle p_i(0)|p_i(T)\rangle e^{-\int_0^T d\eta \langle p_i(\eta)|\dot p_i(\eta)\rangle}~. \label{eq:gprS}
\end{equation}
As shown by the calculation below in Eq.~\eqref{eq:eigva}, we have $p_2(0)=0$. Thus, only the eigenstate corresponding to $p_1(\eta)$ contributes to $\beta$. Now, substituting from Eq.~\eqref{eq:eigva} and \eqref{eq:eigve}, we obtain
\begin{equation}
    \beta= -\omega\int_0^T \cos^2\frac{\lambda(\eta)}{2} d\eta= -\omega\int_0^T \frac{\chi(\eta)- \rho_3(\eta)}{2\chi(\eta)} d\eta~.
\end{equation}
Substituting $\chi(\eta)$ and $\rho_3(\eta)$, one obtains
\begin{multline}
    \beta= -\frac{\omega}{2} \\
    \int\left(1-\frac{-\frac{B}{A} e^{4 A \eta}+\frac{B}{A}+\cos\theta}{\sqrt{\left(-\frac{B}{A} e^{4 A \eta}+\frac{B}{A}+\cos\theta\right)^2+e^{4 A \eta} \sin ^2\theta}}\right) d\eta~.
\end{multline}
Further substitution of $Q= R+ \cos\theta$ and $R= B/A$ reduces this integral to
\begin{equation}
    \beta= -\frac{\omega}{2}
    \int\left(1-\frac{-R e^{4 A \eta}+Q}{\sqrt{\left(-R e^{4 A \eta}+ Q\right)^2+e^{4 A \eta} \sin ^2\theta}}\right) dt~.
\end{equation}
We can express this integral into the standard form by substituting $2 Q R- \sin^2\theta= Q^2+ R^2-1$
\begin{multline}
    \beta= -\frac{\omega}{2}\\
    \int\left(1-\frac{-R e^{4 A \eta}+Q}{\sqrt{R^2 e^{8 A \eta}+ Q^2+e^{4 A \eta} (1-Q^2-R^2)}}\right) d\eta~.
\end{multline}
The solution of this integral is
\begin{equation}
    \beta\approx F(2\pi)-F(0)~, \label{eq:gpex}
\end{equation}
where
\begin{equation}
    \begin{aligned}
    F(\phi)=& - \frac{1}{8 A} \ln\left(\frac{1- Q^2-R^2+2R^2 e^{4 A\phi/\omega}}{2R}+ {\cal S}\right)- \\
    \frac{\text{sgn}(Q)}{8 A}& \ln\left(1- Q^2-R^2+2 \left(Q^2 +|Q| {\cal S} \right) e^{-4 A\phi/\omega}\right) -\frac{\phi}{2}~,\\
    {\cal S}=& \sqrt{R^2 e^{8 A\phi/\omega}+ (1- Q^2-R^2) e^{4 A\phi/\omega}+ Q^2}~.
    \end{aligned}
\end{equation}


\section{$\Lambda$-system restricted to the FP-light cones} \label{app:a}

We obtain an expression for the ground state population when the $\Lambda$-system is restricted to the FP-light cones of Minkowski spacetime. This scenario is equivalent to placing the detector in contact with some thermal bath, where the system’s evolution is governed by the Lindblad master equation
\begin{equation}
    \frac{\partial\rho}{\partial \eta}= -\frac{i}{\hbar}\left[H_\text{eff}, \rho\right]+ \sum_{i,j} \gamma_{ij}(\omega) \left(S_j\rho(\eta)S_i^\dagger- \frac{1}{2}\{S_i^\dagger S_j, \rho(\eta)\} \right)~, \label{eq:linde0}
\end{equation}
where $H_{\text{eff}}= H_0+ H_L$, with $H_{L}= i\hbar \left(K(-\omega)-K(\omega)\right)/2$ representing the Lamb shift Hamiltonian arising from the interaction between the detector and the field. The function $K(\omega)$ is given by
\begin{equation}
     K(\omega)= P \frac{1}{\pi i} \int_{-\infty}^\infty d\alpha \frac{{\cal P}(\alpha)}{\alpha-\omega}~.
\end{equation}
where $P$ denotes the Cauchy principal value of the integral. The matrix $\gamma_{i j}$ encodes the environmental effects, and $S_i$ are the system’s transition operators. Throughout this work, we operate in the regime $\Gamma(\omega)/\omega\ll 1$, where the contribution from the Lamb shift term is negligible.

For our $\Lambda$-system, transitions occur only between the excited state $|e\rangle$ and the ground states $|g_i\rangle, \quad i=1,2$. Since the transition probabilities from $|e\rangle$ to $|g_1\rangle$ and $|e\rangle$ to $|g_2\rangle$ are equal, we denote the corresponding rate by $\gamma(\omega)$. Similarly, for the reverse transitions $|g_1\rangle \to |e\rangle$ and $|g_2\rangle \to |e\rangle$, we use the rate $\gamma(-\omega)$. Expanding Eq.~\eqref{eq:linde0} and using the relation $|\pm\rangle= \left(|g_1\rangle\pm |g_2\rangle\right)/\sqrt{2}$, we obtain
\begin{multline}
    \frac{\partial\rho}{\partial \eta}= -\frac{i}{\hbar}\left[H_0, \rho\right]+ {\cal P}(\omega) {\cal L}[\sigma_{e+}]- {\cal P}(-\omega) {\cal L}[\sigma_{e+}^\dagger] ~, \label{eq:linde1}
\end{multline}
where we have defined $\gamma(\omega)= \sqrt{2} {\cal P}(\omega)$ and ${\cal L}[O]= O \rho O^\dagger - \{O^\dagger O, \rho\}/2$. In component form, this equation is expressed as follows
\begin{equation}
    \begin{aligned}
        \frac{d\rho_{--}}{d\eta}&= 0,\\
    \frac{d\rho_{-+}}{d\eta}&= -i \delta\omega \rho_{-+}- \rho_{-+} (A-B),\\
    \frac{d\rho_{-e}}{d\eta}&= -i \omega \rho_{-e}- \rho_{-e} (A+B),\\
    \frac{d\rho_{++}}{d\eta}&= 2 \rho_{ee} (A+B)- 2 \rho_{++} (A-B),\\
    \frac{d\rho_{+e}}{d\eta}&= -i \omega \rho_{+e}- 2 \rho_{+e} A,\\
    \frac{d\rho_{ee}}{d\eta}&= -2 \rho_{ee} (A+B)+ 2 \rho_{++} (A-B),
    \end{aligned}
\end{equation}
where

\begin{equation}
    A= \frac{1}{4} \left({\cal P}(\omega)+ {\cal P}(-\omega)\right), \quad B= \frac{1}{4} \left({\cal P}(\omega)- {\cal P}(-\omega)\right)~.
\end{equation}
By substituting the value of ${\cal P}(\omega)$ from Eq.~\eqref{eq:drptS}, we obtain
\begin{equation}
    \begin{aligned}
        A=& \frac{\Gamma}{4} \left(1+ \frac{a^2}{\omega^2}\right) \frac{e^{2\pi\omega/a}+1}{e^{2\pi\omega/a}-1}~,\\
    B=& \frac{\Gamma}{4} \left(1+ \frac{a^2}{\omega^2}\right)~.
    \end{aligned}
\end{equation}
with the spontaneous emission rate being $\Gamma= (\omega^3 q^2 |\langle e|r|+\rangle|^2)/(4\pi\epsilon_0\hbar c^3)$. 

Thus, the dynamics has been decomposed into two distinct components: one associated with the $|-\rangle$ state and another corresponding to an effective two-level system involving the $|+\rangle$ and $|e\rangle$ states, which evolves independently. The density matrix corresponding to the initial state of $|\psi(0)\rangle= \sin(\vartheta/2) \cos(\theta/2) |e\rangle+ \sin(\vartheta/2) \sin(\theta/2)|+\rangle + \cos(\vartheta/2) |-\rangle$ is
\begin{widetext}
    \begin{equation}
    \rho(\eta)= \begin{pmatrix}
        \sin^2\frac{\vartheta}{2} e^{-2 A \eta} f(\eta) & \frac{1}{2} e^{-i\omega \eta} e^{-2 A \eta} \sin\theta \sin^2\frac{\vartheta}{2} & \frac{1}{2} e^{-i\omega \eta} e^{-(A+B) \eta} \sin\vartheta \cos\frac{\theta}{2} \\ \frac{1}{2} e^{i\omega \eta} e^{-2 A \eta} \sin\theta \sin^2\frac{\vartheta}{2} & \sin^2\frac{\vartheta}{2}-\sin^2\frac{\vartheta}{2} e^{-2 A \eta} f(\eta) & \frac{1}{2} e^{-(A-B) \eta} \sin\vartheta \sin\frac{\theta}{2} \\
        \frac{1}{2} e^{i\omega \eta} e^{-(A+B) \eta} \sin\vartheta \cos\frac{\theta}{2} & \frac{1}{2} e^{-(A-B) \eta} \sin\vartheta \sin\frac{\theta}{2} & \cos^2\frac{\vartheta}{2}
    \end{pmatrix}~. \label{eq:dmat}
\end{equation}
\end{widetext}
where
\begin{equation}
    f(\eta)= e^{-2 A \eta} \cos^2\frac{\theta}{2}- \left(\frac{B}{A}- 1\right) \sinh(2A \eta).
\end{equation}
One of the eigenvalues of this density matrix is zero, and the corresponding eigenstate defines the dark subspace, which does not acquire a geometric phase during the evolution. The remaining two eigenvalues, associated with the bright subspace, are given by
\begin{equation}
    p_1(\eta)= \frac{1}{2} \left(1+ \chi(\eta)\right), \quad p_2(\eta)= \frac{1}{2} \left(1- \chi(\eta)\right)~, \label{eq:eigva}
\end{equation}
where $\chi(\eta)= \sqrt{\rho_3^2+ e^{-4 A \eta}\sin^2\theta}$ and $\rho_3= e^{-4 A \eta} \cos\theta+ (B/A) \left(e^{-4 A \eta}-1\right)$. The corresponding eigenstates are
\begin{equation}
    \begin{aligned}
        |p_1(\eta)\rangle=& \sin\frac{\lambda(\eta)}{2} |e\rangle+ e^{i\omega \eta} \cos\frac{\lambda(\eta)}{2} |+\rangle~,\\
        |p_2(\eta)\rangle=& \cos\frac{\lambda(\eta)}{2} |e\rangle- e^{i\omega \eta} \sin\frac{\lambda(\eta)}{2} |+\rangle~, \label{eq:eigve}
    \end{aligned}
\end{equation}
where
\begin{equation}
    \tan\frac{\lambda(\eta)}{2}= \sqrt{\frac{\chi(\eta)+ \rho_3(\eta)}{\chi(\eta)- \rho_3(\eta)}}~. \label{eq:iden}
\end{equation}
We can now calculate the ground state population using the relation
\begin{multline}
    P_1= \langle g_1|\rho|g_1\rangle= \rho_{++} |\langle g_1|+\rangle|^2+ \rho_{+-} \langle g_1|+\rangle \langle -|g_1\rangle e^{i(\beta+\varphi)}\\
    + \rho_{-+} \langle g_1|-\rangle \langle +| g_1\rangle e^{-i(\beta+\varphi)}+ \rho_{--} |\langle g_1|-\rangle|^2~,
\end{multline}
where, $\varphi= \Delta{\cal E}\Delta t$ is an initial phase shift that can be tuned by lifting the energy degeneracy of $|+\rangle$ and $|-\rangle$ ($\Delta{\cal E}$) for a short period of time ($\Delta t$). Here, the additional phase factor $e^{i(\beta+\varphi)}$ arises because the bright and dark states of the system evolves in the field as $e^{i(\beta(\eta)+\alpha_b(\eta)+\varphi)}\sqrt{p_1(\eta)+p_2(\eta)}|\psi_b(\eta)\rangle$ and $e^{i\alpha_g(\eta)}\sqrt{p_-}|-\rangle$, respectively. Furthermore, we have assumed that the dynamical phase $\alpha(\eta)$ acquired by the system vanishes during its cyclic evolution, since it is a path-dependent quantity. Consequently, the ground state population simplifies to
\begin{multline}
    P_1(\eta)= \frac{1}{2}\left(1 -\sin^2\frac{\vartheta}{2} e^{-2 A \eta} f(\eta) \right)\\
    + \frac{1}{2} e^{-(A-B) \eta} \sin\vartheta \sin\frac{\theta}{2} \cos\left(\beta(\eta)+\varphi\right)~.
\end{multline}

Now, we express the density matrix of Eq.~\eqref{eq:dmat} in the Bloch-Redfield form~\cite{krantz2019}
\begin{widetext}
    \begin{equation}
    \rho(\eta)= \begin{pmatrix}
        \sin^2\frac{\vartheta}{2} \cos^2\frac{\theta}{2} e^{-\Gamma_1 \eta} & \frac{1}{2} e^{-i\omega \eta} e^{-\Gamma_2 \eta} \sin^2\frac{\vartheta}{2} \sin\theta & \frac{1}{2} e^{-i\omega \eta} e^{-\Gamma_1 \eta/2} \sin\vartheta \cos\frac{\theta}{2} \\ \frac{1}{2} e^{i\omega \eta} e^{-\Gamma_2 \eta} \sin^2\frac{\vartheta}{2} \sin\theta & \sin^2\frac{\vartheta}{2}-\sin^2\frac{\vartheta}{2} \cos^2\frac{\theta}{2} e^{-\Gamma_1 \eta} & \frac{1}{2} \sin\vartheta \sin\frac{\theta}{2} \\
        \frac{1}{2} e^{i\omega \eta} e^{-\Gamma_1 \eta/2} \sin\vartheta \cos\frac{\theta}{2} & \frac{1}{2} \sin\vartheta \sin\frac{\theta}{2} & \cos^2\frac{\vartheta}{2}
    \end{pmatrix}~.
\end{equation}
\end{widetext}
We thus have the density matrix in terms of the longitudional ($\Gamma_1$) and transverse ($\Gamma_2$) relaxation rates. The dephasing process is the combination of the depolarization and the pure dephasing processes
\begin{equation}
    \Gamma_2= \frac{\Gamma_1}{2}+ \Gamma_\phi~.
\end{equation}
We considered the regime where $A \approx B$, which is valid in the low-temperature limit, consistent with the conditions of our analysis. Given the frequencies we are considering, we are operating at the effectively zero temperature regime- this justifies taking $A \approx B$. In this limit, the excitation rate is exponentially suppressed, and the depolarization and dephasing rates are
\begin{equation}
    \Gamma_2= \frac{\Gamma_1}{2}= 2 A= \frac{{\cal P}(\omega)}{2}~. \label{eq:dec}
\end{equation}

\section{Modeling noises}


To account for pure dephasing in fluxonium circuits, which lose coherence through various channels such as dielectric loss and flux noise, we add the dephasing term $\gamma_{k}^\phi {\cal L}[D_k]/2$ to the Lindblad equation~\eqref{eq:linde1}. Here, $D_k = \sum_j M(j,k)\sigma_{jj}$, $M(j,k)=-1$ for $j=k$ and $M(j,k)=1$ otherwise, represents a set of operators modeling the dephasing process, and $\gamma_{k}^\phi$ denotes the pure dephasing rate~\cite{emanuele2018}. Dephasing of bright components, arising partly from longitudinal relaxation (${\cal P}(\omega)$), is already captured by the other terms in Eq.~\eqref{eq:linde1}. Additional dephasing channels therefore only contribute further to this loss of coherence. However, the decoherence of the dark state is entirely unaccounted for by these terms, leaving its impact unknown. To address this, we take a nonzero $\gamma_{-}^\phi$ to model dark-state dephasing and solve the resulting dynamics. 

With the inclusion of $\gamma_{-}^\phi$ term, the Lindblad equation~\eqref{eq:linde1} (written in components) becomes
\begin{equation}
    \begin{aligned}
        \dot\rho_{++}=& \Gamma_1 \rho_{ee}, \quad 
    \dot\rho_{+-}= \left(i\delta\omega- \gamma_-^\phi \right) \rho_{+-}~, \\
    \dot\rho_{-e}=& -\left(i\omega+ \frac{\Gamma_1+2\gamma_-^\phi}{2} \right) \rho_{-e}~, \\
    \dot\rho_{+e}=& -\left(i\omega+ \Gamma_2 \right) \rho_{+e},\quad \dot\rho_{ee}= -\Gamma_1 \rho_{ee}~,
    \end{aligned}
\end{equation}
where $\delta\omega$ is the is the energy of the dark state which is infinitesimally small compared to the energy of the excited state $\omega$. For the initial state $|\psi(0)\rangle= \sin(\vartheta/2) \cos(\theta/2) |e\rangle+ \sin(\vartheta/2) \sin(\theta/2)|+\rangle + \cos(\vartheta/2) |-\rangle$, the density matrix can be written as
\begin{widetext}
    \begin{equation}
    \rho(\eta)= \begin{pmatrix}
        \sin^2\frac{\vartheta}{2} \cos^2\frac{\theta}{2} e^{-\Gamma_1 \eta} & \frac{1}{2} e^{-i\omega \eta} e^{-\Gamma_2 \eta} \sin^2\frac{\vartheta}{2} \sin\theta & \frac{1}{2} e^{-i\omega \eta} e^{-(\Gamma_1+2\gamma_-^\phi) \eta/2} \sin\vartheta \cos\frac{\theta}{2} \\ \frac{1}{2} e^{i\omega \eta} e^{-\Gamma_2 \eta} \sin^2\frac{\vartheta}{2} \sin\theta & \sin^2\frac{\vartheta}{2}-\sin^2\frac{\vartheta}{2} \cos^2\frac{\theta}{2} e^{-\Gamma_1 \eta} & \frac{1}{2} e^{-i\delta\omega \eta} e^{-\gamma_-^\phi \eta} \sin\vartheta \sin\frac{\theta}{2} \\
        \frac{1}{2} e^{i\omega \eta} e^{-(\Gamma_1+2\gamma_-^\phi) \eta/2} \sin\vartheta \cos\frac{\theta}{2} & \frac{1}{2} e^{i\delta\omega \eta} e^{-\gamma_-^\phi \eta} \sin\vartheta \sin\frac{\theta}{2} & \cos^2\frac{\vartheta}{2}
    \end{pmatrix}~.
\end{equation}
\end{widetext}
The geometric phase acquired by the bright state remains unaffected, up to the subleading order in the expansion of $\beta(\eta)$, even when a nonzero dephasing rate $\gamma_{-}^\phi$ is introduced. Following the procedure outlined above, the ground-state population is given by
\begin{multline}
    P_1= \frac{1}{2}\left(1 -\sin^2\frac{\vartheta}{2} \cos^2\frac{\theta}{2} e^{-\Gamma_1 \eta}\right)\\
    + \frac{1}{2} e^{-\gamma_-^\phi \eta} \sin\vartheta \sin\frac{\theta}{2} \cos(\beta+\varphi+\delta\omega \eta)~.
\end{multline}
From this solution, we compute the change in ground state population as the system evolves in a thermal bath, setting $\varphi=0$. We assume a dephasing rate of the dark state $\gamma_-^\phi$ to be approximately $10^4$, consistent with the highest estimate reported in Ref.~\cite{nguyen2019} for a fluxonium device operating at its sweet spot. The maximum energy difference between the bright and dark states is $\delta\omega = 0.06$ GHz, which corresponds to Fig.~2b. Under these parameter values, which are intended to capture the maximal effect of dephasing, the difference in the change in ground state population with and without dephasing is found to be less than one percent.

\section{Circuit parameters}

We fix the asymmetry parameter to be $d=0.13$ and the charging energy to be $E_C= 0.55$ GHz. The Josephson energy $E_J(\phi_1)$ has its maximum value of $E_{J\Sigma}$ when $\phi_1=0$. For our circuit, $E_J(\phi_1)/E_C\gg 1$ is desirable as this condition ensures that the tunneling probability between low energy states tend to vanish, thereby localizing those states inside a well. Therefore, we choose $E_{J\Sigma}/E_C=50$, so that $E_{J\Sigma}= 27.5$ GHz, a value realized in the literature. The energy levels of this circuit constitute a $\Lambda$-system that we need for our experiment over some range of $E_J(\phi_1)$ if we take $E_L= 0.72$ GHz. Also, to maintain the $\Lambda$-system, we can only tune the energy $E_J(\phi_1)$ such that the phase $\phi_1$ is confined to the range $(2.50,2.80)$. The phase $\phi_2$ should be adjusted accordingly such that two local minima lying withing the range $(-\pi,\pi)$ of the potential well, corresponding to the Hamiltonian of Eq.~(1), remain symmetric. We performed approximate analytical calculations to show that this ``sweet spot" occurs at $\phi_2= \pi-\phi_J(\phi_{1})$, which agrees well with our numerical results. The analytical approach involves first determining the values of $\phi$ corresponding to the minimum potential, which numerical simulations indicate are located near $\pi$ and $-\pi$. We then equate the expression for potential at these values of $\phi$ to derive the dependence of $\phi_2$ on $\phi_1$.

\begin{figure}[!tbp]
    \centering
    \includegraphics[width=0.48\textwidth]{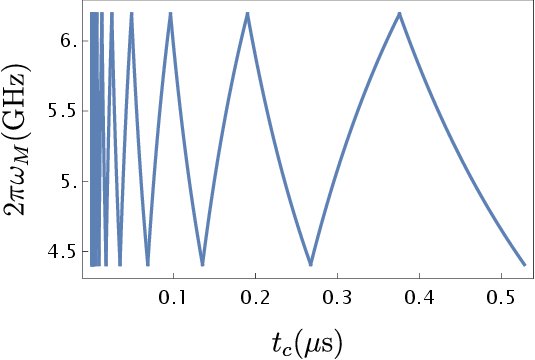}
    \caption{
    The tuning profile of the energy levels of the circuit during $530$ ns evolutions in Minkowski time (equivalent to approximately $25$-quasicycle evolutions in FP time) is achieved by modulating the external fluxes. This modulation, where energy levels of a detector scale inversely with Minkowski time corresponds to a detector at rest in FP coordinates. As a result, geometric phase change is induced in the detector, which is encoded in its ground state population. 
    }
    \label{fig:s4}
\end{figure}

With these parameter values, the modified fluxonium circuit acts as a tunable $\Lambda$-system, whose energy levels can be tuned in the range $\omega^i_M/2\pi= 4.4$ GHz to $\omega^f_M/2\pi= 6.2$ GHz. The change in ground state population as the energy levels of the $\Lambda$-detector is modulated between $\omega^i_M/2\pi$ and $\omega^f_M/2\pi$ is shown in Fig.~3. The energy levels undergo $25$-quasicycles evolution in $1\mu s$, where each quasicycle corresponds to an evolution in conformal time from $\eta=0$ to $\eta= 2\pi/\omega$. The tuning of the circuit's frequency in Minkowski time can be calculated using Eq.~(17), which is illustrated in Fig.~\ref{fig:s4}. It is important to note that the sharp edges observed in the graph are not physically realizable in experimental settings. However, smoothly tuning the energy profile in practice is unlikely to significantly alter our results, as smoothing the edges affects a narrow region within the circuit’s tuning window.

\putbib[origin1]
\end{bibunit}

\end{document}